\documentclass[12pt,a4paper]{article}
\usepackage[
        left=3cm,
        right=2cm,
        top=3cm,
        bottom=2cm,
]{geometry}
\usepackage{amsfonts,amsmath,amssymb}
\usepackage[pdftex]{graphicx}
\usepackage{latexsym}
\usepackage{psfrag}
\usepackage{authblk}
\usepackage{xcolor}
\usepackage{bm}
\usepackage{lineno}
\usepackage{cite}
\usepackage{setspace}
\usepackage{doi,verbatim}
\usepackage{hyperref}
\usepackage{subfig}
\usepackage{soul}
\begin{document}
%
%\doublespacing

\title{ Modulated structures in a Lebwohl-Lasher model with chiral interactions }
\author[1]{E. S. Nascimento\thanks{edusantos18@esp.puc-rio.br}}
\author[2]{A. Petri\thanks{alberto.petri@isc.cnr.it}}
\author[3]{S. R. Salinas\thanks{ssalinas@if.usp.br}} 
\affil[1]{ Depto de F\'{\i}sica, PUC-Rio, Rio de Janeiro, RJ, Brazil }
\affil[2]{ CNR - Istituto dei Sistemi Complessi, Dipartimento di Fisica, Universit\`{a} Sapienza, Roma, Italy }            
\affil[3]{ Instituto de F\'{\i}sica, USP, S\~{a}o Paulo, SP, Brazil }            

\date{}

\maketitle

\begin{abstract}
We consider a Lebwohl-Lasher lattice model with nematic directors restricted
to point along $p$ planar directions. This $XY$ Lebwohl-Lasher system is the
nematic analogue of the standard $p$-state clock model. We then include chiral
interactions, and introduce a chiral $p$-state nematic clock model. The
statistical problem is formulated as a discrete non-linear map on a Cayley
tree. The attractors of this map correspond to the physical solutions deep in
the interior of the tree. It is possible to observe uniform and periodic
structures, depending on temperature and a parameter of chirality. We find
many different chiral nematic phases, and point out the effects of temperature
and chirality on the modulation associated with these structures.
\end{abstract}


\section{Introduction}

Competing interactions are the basic mechanism to describe the onset of
sequences of modulated phases in many physical systems
\cite{Seul1995,Andelman2009,Giuliani2009,SelkeBook}. In magnetism, the most
investigated lattice statistical model to account for the existence of
spatially modulated structures is the ANNNI model
\cite{Bak82,Selke88,Yeomans88}, which is an Ising system with competing ferro
and antiferromagnetic interactions between first and second-neighbor spins
along an axial direction. The ANNNI model exhibits one of the richest phase
diagrams in the literature, with a wealth of modulated structures depending on
temperature and on a parameter associated with the ratio between the strengths
of the competing couplings. It is remarkable that the modulation associated
with these periodic structures displays a nontrivial, staircase-like behavior,
as a function of the thermodynamic field parameters \cite{Nascimento2014}.

There is an alternative mechanism of competition, which is suggested by an
earlier proposal of Dzyaloshinskii and Moriya to explain the modulated
structures of helimagnets \cite{DM1,DM2,DM3}. According to this Dzyaloshinskii
- Moriya (DM) mechanism, in addition to the standard exchange interactions,
first-neighbor vector spins along an axis are supposed to interact via chiral
couplings. In classical statistical mechanics, it is possible to show that the DM 
formulation is a generalization of the $p$-state chiral clock (CC) models  
introduced by Huse \cite{Huse81} and Ostlund \cite{Ostlund1981}, with planar vector spin
variables along $p$ directions, and the inclusion of chiral interactions
between first-neighbors along an axis. These CC systems have been shown to lead
to similar complex modulated structures as the ANNNI model. The phase diagrams
of the CC models display sequences of many different helical ferromagnetic
phases, and a complex behavior of the main wave numbers in terms of
temperature \cite{Huse81,Ostlund1981,Ottinger83,Siegert85,Pleimling98}.

Spatially modulated structures with helical ordering are not exclusive of
magnetic systems. In soft-matter physics, simple cholesteric phases in
liquid-crystalline compounds provide perhaps the best examples of chiral
nematic structures, in which the nematic director exhibits a spatial variation
of helical type along a given direction \cite{DeGennes}. Many physical aspects
of cholesteric nematics can be studied in the framework of the
phenomenological approaches, along the lines of the Landau-de Gennes theory,
with the inclusion of adequate terms associated with the spatial variation of
the nematic director \cite{Kamien2001,Selinger2016}.

From the molecular point of view, according to an early work by Goossens
\cite{Goossens1971}, the consideration of an induced dipole-quadrupole
contribution to the dispersion interactions can give rise to a cholesteric
phase. This picture has been extended by some investigators
\cite{vanderMeer1976} \cite{Krutzen1989}, with the formulation of a
statistical model, of mean-field type, which might be able to describe the
temperature effects in a cholesteric phase. In the calculations of van der
Meer and collaborators \cite{vanderMeer1976}, there is a modulated structure,
which is not affected by temperature. Although there is no attempt to draw a
global phase diagram, the numerical mean-field calculations of Krutzen and
Vertogen \cite{Krutzen1989} point out the possibility of a
temperature-dependent pitch if one includes a nematic-twist term in the
original pair potential. Along Goossens's ideas, Lin-Liu and collaborators
\cite{Lin-Liu1976} considered a planar model for the cholesteric phase, which
does lead to the description of a temperature sensitive pitch. In these
earlier statistical mechanics calculations it is difficult to point out the
effects of chirality. Also, there is no attempt to draw a global phase diagram
in terms of temperature and a parameter to gauge the strength of chiral
interactions. In fact, by means of these earlier calculations, it is not 
clear to see how thermal fluctuations affect the cholesteric pitch.

The statistical mechanics calculations for the magnetic phase diagrams of the
ANNNI and CC models provided the inspiration to carry out the present work.
Similar calculations were still missing for the Dzyaloshinskii-Moriya picture
of model systems with nematic interactions. We then formulated a $p$-state
CC model with head-tail symmetry, and checked the existence of
long-period structures in a nematic setting. The main goal of this work is the
establishment of qualitative contacts with the description of cholesteric
structures in liquid crystals. In this preliminary analysis, calculations have
been restricted to the simplest versions of these nematic clock models, which
are already sufficient to point out the statistical origin of spatially
modulated phases, to draw representative global phase diagrams, and give an
indication of the change of the pitch with temperature and model parameters.
We remark that it is easy to see that there is a correspondence between the
essential terms in the Hamiltonians associated with the $p$-state nematic clock model 
and of a planar version of Goossen's proposal, as it has been used by Krutzen 
and Vertogen \cite{Krutzen1989} and by Lin-Liu and collaborators \cite{Lin-Liu1976}.

This paper is organized as follows. In Section \ref{XYLL}, we consider a version of the
Lebwohl-Lasher (LL) model \cite{Lebwohl72}, with the restriction of the nematic directors to
point along a discrete set of directions on the $x-y$ planes. On the basis of
this LL model, we define a $p$-state nematic clock model (NC), which is the
nematic counterpart of the standard (magnetic) clock models. We then introduce
chiral interactions along the axis normal to the $x-y$ planes. In Section \ref{6CLLL}, 
we define a $p$-state chiral nematic clock model, which we call CNC model.
Also, instead of performing a laborious layer-by-layer mean-field calculation, we
formulate the problem on a Cayley tree, and restrict the analysis to the
simplest, and nontrivial, chiral nematic model, with just $p=6$ angular
directions. The problem is reduced to the investigation of a discrete
nonlinear map, whose attractors correspond to physical solutions in the deep
interior of large Cayley tree (which is known as a Bethe lattice). It should
be remarked that we take full advantage of the geometry of a Cayley tree,
which is particularly adequate to describe modulation effects along a radial
direction, as it has been pointed out in previous work for the ANNNI and
chiral clock models. In Section \ref{MFpCLLL}, we further simplify the problem by
considering the infinite-coordination (mean-field) limit of a Cayley tree,
which is a further simplification, and which is sufficient to illustrate the
main features of a chiral nematic model system. We show that the global phase
diagram of these systems displays many different chiral nematic phases, and
give an illustration to show that these modulated structures exhibit a complex
behavior in terms of model parameters. Some conclusions as well an outlook of
future work are presented in Section \ref{Con}.

\section{The chiral nematic clock model} \label{XYLL}

We consider pair interactions between nematogenic molecules on the sites of
a cubic lattice. In the context of the Maier-Saupe approach to the nematic
phase transitions \cite{OLC2017,DeGennes}, we write the Hamiltonian
\begin{equation}
\mathcal{H}_{MS}=-J\sum_{\left( i,j\right) }\sum_{\mu ,\nu }Q_{i}^{\mu
\nu }\,Q_{j}^{\mu \nu }=-J\sum_{\left( i,j\right) }\mathbf{Q}%
_{i}\cdot \mathbf{Q}_{j},
\end{equation}
where $J>0$ is a positive energy parameter, the first sum is over pairs of
lattice sites, $\mu ,\nu $ are Cartesian coordinates, and $\mathbf{Q}_{i}$
is a local traceless quadrupolar tensor.

We now restrict the microscopic nematic directors to the $x-y$ plane. In
order to account for the head-tail symmetry of nematic liquid crystals, we
write a $2\times 2$ traceless tensor,%
\begin{equation}
Q_{\mu \nu }=\frac{1}{2}\left( 2n_{\mu }n_{\nu }-\delta _{\mu \nu }\right) ,
\end{equation}%
where $\mu ,\nu $ are restricted to $x,y$, $\delta _{\mu \nu }$ is the usual
Kronecker symbol, and $n_{\mu },n_{\nu }$ are the components of a
two-dimensional microscopic nematic director,%
\begin{equation}
\overrightarrow{n}=\left( 
\begin{array}{c}
\cos \theta \\ 
\sin \theta%
\end{array}%
\right) .
\end{equation}%
We then have the traceless tensor%
\begin{equation}
\mathbf{Q}=\frac{1}{2}\left( 
\begin{array}{cc}
2\cos ^{2}\theta -1 & 2\cos \theta \sin \theta \\ 
2\cos \theta \sin \theta & 2\sin ^{2}\theta -1%
\end{array}%
\right) ,  \label{Q}
\end{equation}%
where $\theta $ is the angle of the (planar) nematic microscopic director
with the $x$ axis.

Given the local director elements, 
\begin{equation}
\overrightarrow{n}_{i}=\left( 
\begin{array}{c}
\cos \theta _{i} \\ 
\sin \theta _{i}%
\end{array}%
\right) ,\qquad \overrightarrow{n}_{j}=\left( 
\begin{array}{c}
\cos \theta _{j} \\ 
\sin \theta _{j}%
\end{array}%
\right) ,
\end{equation}%
and discarding a harmless constant term, it is straightforward to write the
Hamiltonian of a nematic clock (NC) model, 
\begin{equation}
\mathcal{H}_{NC}=-J\sum_{\left( i,j\right) }\cos ^{2}\left( \theta
_{i}-\theta _{j}\right) ,  \label{LLL}
\end{equation}%
which is an $XY$ version of the well known Lebwohl-Lasher model (and a
nematic analogue of the classical $XY$ spin model of magnetism).

We now assume that chiral effects are mimicked by a rotation of the nematic
directors around the $z$ axis (which is the normal direction to the $x-y$
planes). We then write%
\begin{equation}
\overrightarrow{n}_{rot}=\left( 
\begin{array}{cc}
\cos \Delta & \sin \Delta \\ 
-\sin \Delta & \cos \Delta%
\end{array}%
\right) \left( 
\begin{array}{c}
\cos \theta \\ 
\sin \theta%
\end{array}%
\right) =\left( 
\begin{array}{c}
\cos \left( \theta -\Delta \right) \\ 
\sin \left( \theta -\Delta \right)%
\end{array}%
\right) ,
\end{equation}%
where the angle $\Delta $ is the parameter that gauges chirality. The
associated tensor is given by%
\begin{equation}
\mathbf{Q}_{rot}=\frac{1}{2}\left( 
\begin{array}{cc}
2\cos ^{2}\left( \theta -\Delta \right) -1 & 2\cos \left( \theta -\Delta
\right) \sin \left( \theta -\Delta \right) \\ 
2\cos \left( \theta -\Delta \right) \sin \left( \theta -\Delta \right) & 
2\sin ^{2}\left( \theta -\Delta \right) -1%
\end{array}%
\right) .
\end{equation}%
Therefore, the pair interaction between nematogenic elements at sites $i$
and $j$ along the $z$ direction can be written as%
\begin{equation}
\mathbf{Q}_{i}\cdot \mathbf{Q}_{rot,j}=\cos ^{2}\left( \theta _{i}-\theta
_{j}+\Delta \right) -\frac{1}{2},
\end{equation}
which leads to a general form of the Hamiltonian of the chiral nematic clock
(CNC) model, 
\begin{equation}
\mathcal{H}_{CNC}=-J_{0}\sideset{}{^\perp}\sum_{\left( i,j\right) }\,\cos ^{2}
\left( \theta _{i}-\theta _{j}\right) -J\sideset{}{^\parallel}\sum_{\left(
i,j\right) }\,\left[ \cos ^{2}\left( \theta _{i}-\theta
_{j}+\Delta \right) -\frac{1}{2}\right] ,  \label{CLLL}
\end{equation}%
where the first sum is over pair interactions along the $x-y$ planes of the
cubic lattice, and the second sum is over pair interactions along the $z$
direction of the lattice. We remark that $J_{0}>0$ and $J_{1}>0$ are energy
parameters, and $0<\Delta <2\pi $ is the angular parameter that gauges the
degree of chirality.

If we consider just $p$ angular directions, this expression is the natural nematic 
extension of the well-known $p$-state chiral clock model \cite{Huse81,Ottinger83}, 
which has been appropriately generalized to describe chiral nematic-like phases. 
Also, it is straightforward to notice that, for $p$ even, a clock system with 
local cosine squared interaction is equivalent to a usual clock model with $p/2$ states. 
As a result, the CNC model with $p$ even presents the same symmetry properties as CC 
models \cite{Huse81,Ostlund1981,Yeomans88}, which are given by the transformations
\begin{equation} \label{Sym}
 \begin{split}
  \Delta &\to \frac{2\pi}{p} - \Delta, \\
  \theta_i &\to \frac{2\pi}{p} - \theta_i.
 \end{split}
\end{equation}
These symmetry relations simplify the analysis of the phase behavior of the model.

\section{Restricted-orientation model on a Cayley tree} \label{6CLLL}

It is important to take into account that the description of macroscopic nematic 
structures requires the assumption of a head-tail symmetry of the microscopic states. 
This assumption suggests that we should restrict the considerations to $p$-state
CC models with even values of the integer $p$. However, the simplest case, with $p=4$, 
does not present complex modulated structures, as we explicitly show in 
Appendix \ref{4state}.

\begin{figure}[h]
\centering
\includegraphics[scale=0.3]{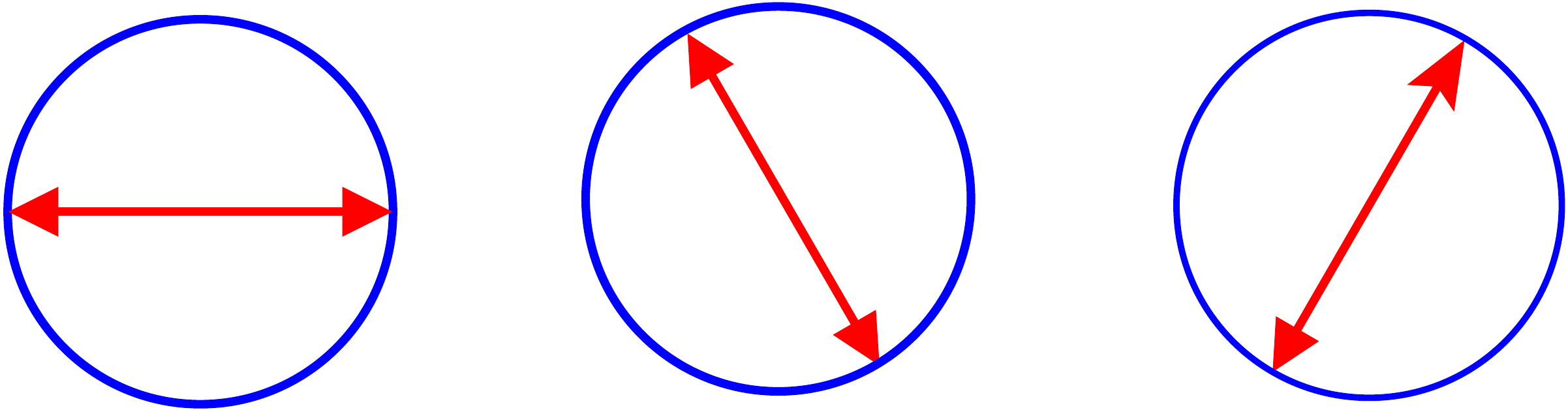}
\caption{States for the planar Lebwohl-Lasher model. The head-tail symmetry effectively maps 
a $6$-state vector variable to a $3$-state ``nematic clock'' variable.}
\label{6states}
\end{figure}

We then consider a $6$-state model, with a discrete choice of
angular variables,%
\begin{equation}
\theta\left(  k\right)  =\frac{\pi}{3}\left(  k-1\right)  ,\quad k=1,...,6.
\end{equation}
Due to the head-tail symmetry of the nematic systems, the pair of angular
variables $\theta\left(  1\right)  $ and $\theta\left(  4\right)  $ leads to a
single tensor variable, which we call $\mathbf{Q}\left(  1\right)  $. Also,
$\theta\left(  2\right)  $ and $\theta\left(  5\right)  $ lead to $\mathbf{Q}\left(
2\right)  $, and the angular variables $\theta\left(  3\right)  $ and
$\theta\left(  6\right)  $ lead to $\mathbf{Q}\left(  3\right)  $. We then use equation
\eqref{Q} to write the microscopic tensor variables,
\begin{equation}%
\begin{split}
\mathbf{Q}\left(  1\right)   &  =\frac{1}{2}%
\begin{pmatrix}
1 & 0\\
0 & -1
\end{pmatrix}
,\\
\mathbf{Q}\left(  2\right)   &  =\frac{1}{4}%
\begin{pmatrix}
-1 & \sqrt{3}\\
\sqrt{3} & 1
\end{pmatrix}
,\\
\mathbf{Q}\left(  3\right)   &  =\frac{1}{4}%
\begin{pmatrix}
-1 & -\sqrt{3}\\
-\sqrt{3} & 1
\end{pmatrix}
.
\end{split}
\label{Q3}%
\end{equation}
In this nematic context, the problem is reduced to the consideration of just
these three local tensor states, as schematically represented in Figure \ref{6states}.

If we assume a set of effective fields along the $z$ direction, it is quite
natural to carry out a layer-by-layer mean-field calculation to obtain the
phase diagrams associated with the complete model Hamiltonian given by \eqref{CLLL}. 
These quite laborious mean-field calculations have indeed been
carried out for analyzing the details of the rich phase diagrams of the
ANNNI and standard chiral clock models \cite{Selke88}. We now turn to a
different and easier approach, which will lead to the same qualitative
features of the phase diagrams. Since we are interested in the main overall
effects of chirality, it is certainly convenient to take advantage of the
radial structure of a Cayley tree. In addition, we perform numerical
calculations for the problem in the limit of infinite coordination of a
Cayley tree, which is expected to present the same qualitative results of
the mean-field calculations.

\begin{figure}
\centering
\includegraphics[scale=0.2]{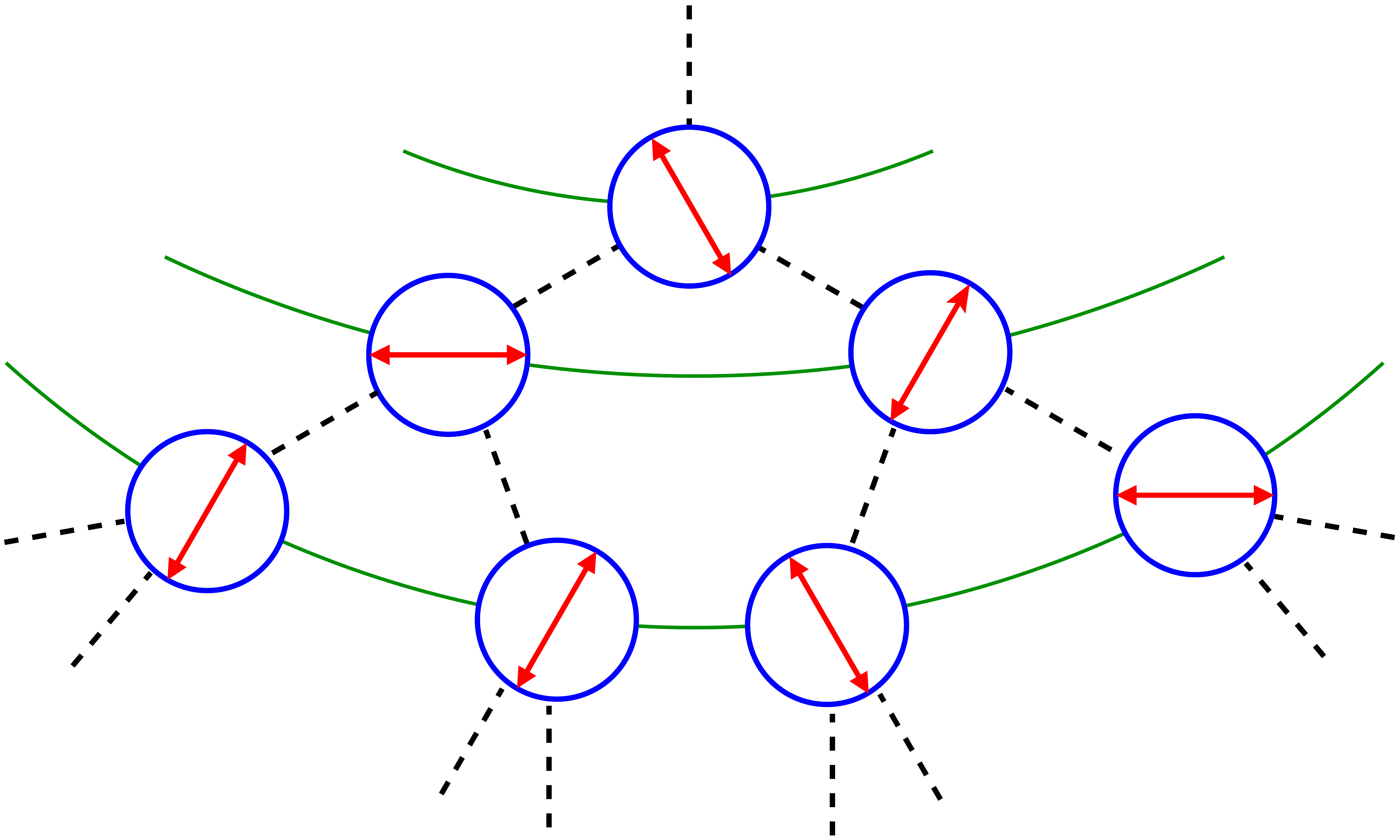}
\caption{Local structure of the planar Lebwohl-Lasher model on a 
Cayley tree with coordination three. ``Nematic clocks'' sit on the vertices of the tree.
Interactions (dashed lines) are between next-neighbors along the branches of the tree.}
\label{CayleyCLL}
\end{figure}

We now restrict equation (\ref{CLLL}) to the interactions along the $z$
direction, and write the Hamiltonian
\begin{equation}
\mathcal{H}=-J\sum_{\left( i,j\right)}\,\left[ \cos
^{2}\left( \theta _{i}-\theta _{j}+\Delta \right) -\frac{1}{2} \right] ,
\end{equation}
for a $6$-state chiral nematic clock model along the branches of a Cayley
tree of ramification $r\geq 1$. Similar calculations on a Cayley tree have
been carried out for a standard $p$-state chiral clock model \cite{Yokoi85}
and for some analogs of the ANNNI model \cite{Oliveira85}. It is well known
that problems on a Cayley tree can be formulated in terms of discrete
non-linear dissipative maps, whose attractors lead to solutions of
physical interest. Along the lines of these earlier calculations on a Cayley
tree, we write the partial partition function of a tree of $n+1$
generations, $Z_{n+1}\left( i\right) $, in the tensor state $\mathbf{Q}%
\left( i\right) $, for $i=1,2,3$, at the root site, in terms of the three
partial partitions functions, $Z_{n}\left( 1\right) $, $Z_{n}\left( 2\right) 
$, and $Z_{n}\left( 2\right) $, of the connected trees of the previous $n$
generation. Figure \ref{CayleyCLL} exhibits schematically the branching of a 
Cayley tree with coordination three, where each vertex presents a nematogenic clock 
interacting with its next-neighbors, along the tree generations. 
Physical solutions, in the deep interior of a large tree,
correspond to the attractors of the non linear map. This deep interior of a
very large tree is known as a Bethe lattice, since the physical solutions
are expected to agree with a simple pair or Bethe-Peierls approximation. As
we have mentioned, in the limit of infinite coordination (and vanishing
interactions), $r\rightarrow \infty $, $J\rightarrow 0$, with $rJ$ fixed, we
are supposed to recover the main features of the standard mean-field results 
\cite{Yokoi85,Oliveira85}.

We now adopt a more compact notation, and write a set of three recurrence
relations between the partial partition functions of a tree of $n+1$
generations, $Z_{1}^{\prime}=Z_{n+1}\left(  1\right)  $, $Z_{2}^{\prime
}=Z_{n+1}\left(  2\right)  $, and $Z_{3}^{\prime}=Z_{n+1}\left(  3\right)  $,
and the partition functions of a tree of $n$ generations. Although it demands
some algebraic effort, it is not difficult to obtain%
\begin{equation}%
\begin{split}
Z_{1}^{\prime}  &  =\left[  A\,Z_{1}+B\,Z_{2}+C\,Z_{3}\right]  ^{r},\\
Z_{2}^{\prime}  &  =\left[  C\,Z_{1}+A\,Z_{2}+B\,Z_{3}\right]  ^{r},\\
Z_{3}^{\prime}  &  =\left[  B\,Z_{1}+C\,Z_{2}+A\,Z_{3}\right]  ^{r},
\end{split}
\label{rec}%
\end{equation}
where
\begin{equation}%
\begin{split}
A  &  =\exp\left[  \beta J\left(  \cos^{2}\Delta-\frac{1}{2}\right)  \right]
,\\
B  &  =\exp\left\{  \beta J\left[  \cos^{2}\left(  \frac{\pi}{3}%
+\Delta\right)  -\frac{1}{2}\right]  \right\}  ,\\
C  &  =\exp\left\{  \beta J\left[  \cos^{2}\left(  \frac{\pi}{3}%
-\Delta\right)  -\frac{1}{2}\right]  \right\}  ,
\end{split}
\label{ABC}%
\end{equation}
and $\beta$ is the inverse of temperature.

It is interesting to rewrite this map in terms of density variables,
\begin{equation}
\rho_{i}=\frac{Z_{i}}{Z_{1}+Z_{2}+Z_{3}},
\end{equation}
for $i=1$, $2$, $3$. Taking into account that%
\begin{equation}
\rho_{1}+\rho_{2}+\rho_{3}=1,
\end{equation}
we further reduce the problem to a two-dimensional map, in terms of two
densities only,%
\begin{equation}%
\begin{split}
\rho_{1}^{\prime} &  =\frac{1}{D}\left[  C+\left(  A-C\right)  \rho
_{1}+\left(  B-C\right)  \rho_{2}\right]  ^{r},\\
\rho_{2}^{\prime} &  =\frac{1}{D}\left[  B+\left(  C-B\right)  \rho
_{1}+\left(  A-B\right)  \rho_{2}\right]  ^{r},
\end{split}
\label{d1d2}%
\end{equation}
with
\begin{equation}%
\begin{split}
D &  =\left[  C+\left(  A-C\right)  \rho_{1}+\left(  B-C\right)  \rho
_{2}\right]  ^{r}+\\
&  \quad+\left[  B+\left(  C-B\right)  \rho_{1}+\left(  A-B\right)  \rho
_{2}\right]  ^{r}+\\
&  \qquad+\left[  A+\left(  B-A\right)  \rho_{1}+\left(  C-A\right)  \rho
_{2}\right]  ^{r},
\end{split}
\label{D}%
\end{equation}
where $A$, $B$, and $C$, are given by \eqref{ABC}. The analysis of this
two-dimensional system of equations leads to the stability borders of the
phase diagrams in terms of temperature and the chiral parameter $\Delta$. It
is easy to see that there is a trivial (disordered) fixed point, $\rho
_{1}^{\ast}=\rho_{2}^{\ast}=1/3$, which is linearly stable at sufficiently
high temperatures.

Instead of working with the densities, we can change to more convenient
variables from the physical point of view. We then consider the average value
of the elements of the symmetric and traceless tensor $\mathbf{Q}$, given by
equation \eqref{Q}, and define two new variables,%
\begin{equation}%
\begin{split}
q_{1} &  =\left\langle Q_{xx}\right\rangle =\left\langle \cos^{2}%
\theta\right\rangle -\frac{1}{2},\\
q_{2} &  =\left\langle Q_{xy}\right\rangle =\frac{1}{2}\left\langle
\sin2\theta\right\rangle .
\end{split}
\end{equation}
Using the densities $\rho_{1}$, $\rho_{2}$, and $\rho_{3}$, we write%
\begin{equation}%
\begin{split}
q_{1} &  =\frac{3}{4}\rho_{1}-\frac{1}{4},\\
q_{2} &  =\frac{\sqrt{3}}{4}\left(  -1+\rho_{1}+2\rho_{2}\right)  ,
\end{split}
\end{equation}
from which we have the densities in terms of new variables, $q_{1}$ and
$q_{2}$,
\begin{equation}%
\begin{split}
\rho_{1} &  =\frac{4}{3}q_{1}+\frac{1}{3},\\
\rho_{2} &  =\frac{1}{3}-\frac{2}{3}q_{1}+\frac{2}{\sqrt{3}}q_{2}.
\end{split}
\end{equation}
Therefore, if we insert these expressions in (\ref{d1d2}), we can as well work
with a two-dimensional map in terms of the alternative variables $q_{1}$ and
$q_{2}$, which are directly related to the tensor order parameter. It is
immediate to see that $q_{1}^{\ast}=q_{2}^{\ast}=0$ corresponds to the
disordered fixed point ($\rho_{1}^{\ast}=\rho_{2}^{\ast}=1/3$) of this model system.

\begin{figure} [h]
\centering
 \subfloat[Nematic state]{\includegraphics[scale=0.7,angle=-90]{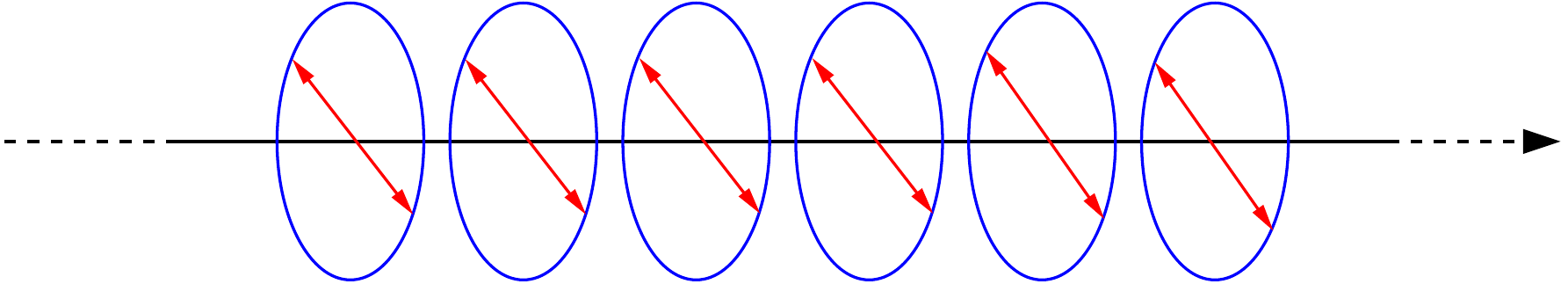}\label{GS:f1}} \\
 \vspace{1cm}
 \subfloat[$1/3$ state]{\includegraphics[scale=0.7,angle=-90]{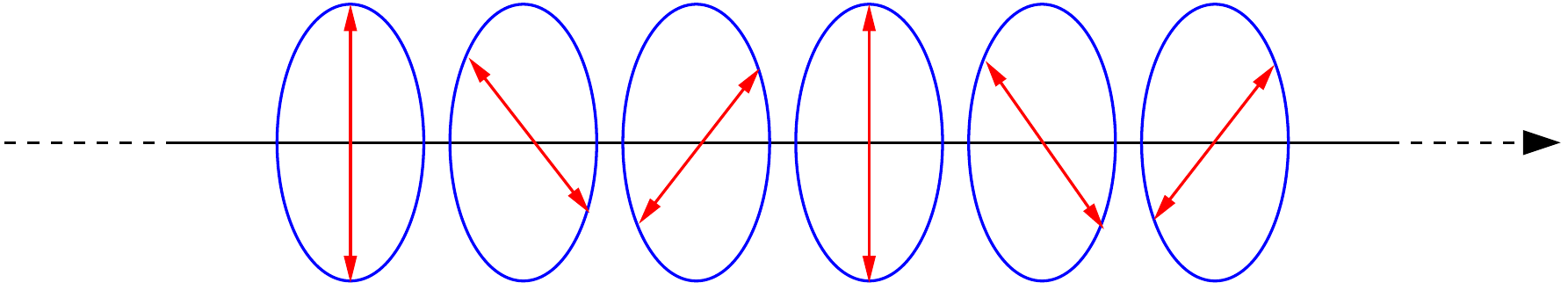}\label{GS:f2}}
\caption{Ground states structures for the chiral nematic clock system. (a) Nematic ordering for $0 \le \Psi < 1/2 $. 
 (b) Chiral, right-handed, ordering with wavenumber $q=2\pi/3$ for $1/2 < \Psi \le 1 $.}
\label{GS}
\end{figure}

It is easy to analyze a number of statistical properties of one-dimensional models with chiral short-range bonds. 
For the particular system we are investigating, with ramification $r=1$, it is possible to show that
\begin{equation}
\left(
\begin{array}
[c]{c}%
Z_{1}^{\prime}\\
Z_{2}^{\prime}\\
Z_{3}^{\prime}%
\end{array}
\right)  =\mathbf{M}\,\left(
\begin{array}
[c]{c}%
Z_{1}\\
Z_{2}\\
Z_{3}%
\end{array}
\right)  ,
\end{equation}
where $\mathbf{M}$ is a cyclic matrix,%
\begin{equation}
\mathbf{M}\,=\left(
\begin{array}
[c]{ccc}%
A & B & C\\
C & A & B\\
B & C & A
\end{array}
\right)  .
\end{equation}
The eigenvalues of this cyclic matrix are given by%
\begin{equation}%
\begin{split}
\Lambda_{0} &  =A+B+C,\\
\Lambda_{\pm} &  =A-\frac{1}{2}\left(  B+C\right)  \pm i\frac{\sqrt{3}}%
{2}\left(  B-C\right)  .
\end{split}
\end{equation}
Note the presence of the complex conjugate eigenvalues $\Lambda_{+}$ and
$\Lambda_{-}$, which lead to an oscillating decay of the pair correlation
function for most values of $\Delta$. This is already an indication
of the existence of spatially modulated structures in larger dimensional
systems. The fixed point, $\rho_{1}^{\ast}=\rho_{2}^{\ast}=1/3$, is linearly
stable, except at zero temperature. 
In addition to the complex structure of correlations for non-zero temperatures, the model also presents a 
rich ground state physics, which is more appropriated to investigate by means of a dimensionless chiral field,
\begin{equation} \label{ChiralField}
 \Psi= \frac{3\Delta}{\pi}.
\end{equation}
As a result, one can notice that the ground state behavior is the nematic analogous of the ferromagnetic 
CC Hamiltonians\cite{Yeomans1982,Yeomans88}. Then, for $0 \le \Psi < 1/2 $, the ground state is of nematic 
type, where all nematic clocks are in the same orientational state, as indicated schematically in Figure \ref{GS:f1}. 
However, for $1/2 < \Psi \le 1 $, the system exhibits a right-handed helical ordering, with modulation 
$q = 2\pi/3$ (period $3$), along the chiral direction (see Figure \ref{GS:f2} ).
The special chiral field value $\Psi=1/2$ leads the system to an infinitely degenerated ground state, 
with finite zero-temperature entropy, usually called multiphase point \cite{Selke88,Yeomans88}. 
We expect, as discussed in the context of ANNNI models \cite{Selke88} and CC spin systems \cite{Yeomans1982}, 
that thermal fluctuations may break the ground state degenerescence, which can give rise to nontrivial
phase behavior for finite temperatures.

\section{Mean-field limit and phase behavior} \label{MFpCLLL}

We now resort to a well known technique to obtain mean-field results from the
recursion relations on a Cayley tree. We take an ``infinite coordination limit'', 
$r\to\infty$ and $J\to0$, while keeping the product $rJ$ fixed. In this limit, the
nonlinear map becomes somewhat more feasible to deal with, which allows us to
grasp the main qualitative features of this problem.

In the infinite coordination limit, equations \eqref{d1d2} can be rewritten
as
\begin{equation} 
\rho_{1}^{\prime}=\frac{\exp\left(  \beta JrM_{1}\right)  }{%
%TCIMACRO{\dsum \limits_{i=1}^{3}}%
%BeginExpansion
{\displaystyle\sum\limits_{i=1}^{3}}
%EndExpansion
\exp\left(  \beta JrM_{i}\right)  },\quad\rho_{2}^{\prime}=\frac{\exp\left(
\beta JrM_{2}\right)  }{%
%TCIMACRO{\dsum \limits_{i=1}^{3}}%
%BeginExpansion
{\displaystyle\sum\limits_{i=1}^{3}}
%EndExpansion
\exp\left(  \beta JrM_{i}\right)  },\label{MF}%
\end{equation}
with%
\begin{equation}%
\begin{split}
M_{1} &  =-\frac{1}{4}\left(  \cos2\Delta+\sqrt{3}\sin2\Delta\right)
+\frac{1}{4}\left(  3\cos2\Delta+\sqrt{3}\sin2\Delta\right)  \rho_{1}+\\
&  \qquad+\frac{\sqrt{3}}{2}\left(  \sin2\Delta\right)  \rho_{2},\\
& \\
M_{2} &  =-\frac{1}{4}\left(  \cos2\Delta-\sqrt{3}\sin2\Delta\right)
-\frac{\sqrt{3}}{2}\left(  \sin2\Delta\right)  \rho_{1}+\\
&  \qquad+\frac{1}{4}\left(  3\cos2\Delta-\sqrt{3}\sin2\Delta\right)  \rho
_{2},\\
& \\
M_{3} &  =\frac{1}{2}\cos2\Delta  -\frac{1}{4}\left(
3\cos2\Delta-\sqrt{3}\sin2\Delta\right)  \rho_{1}-\\
&  \qquad-\frac{1}{4}\left(  3\cos2\Delta+\sqrt{3}\sin2\Delta\right)  \rho
_{2}.
\end{split}
\end{equation}
The system of coupled nonlinear equations \eqref{MF} can be solved iteratively, and the solutions 
allow us to draw the global phase diagram as well the behavior of the modulation as a function of 
model parameters.

It is immediate to show that the mean-field equations \eqref{MF} lead to a disordered
fixed point,
\begin{equation}
 \rho_{1}^{\ast}=\rho_{2}^{\ast}=\frac{1}{3},
\end{equation}
for all values of temperature and chiral parameter. A linear analysis of
stability indicates that this disordered solution is unstable below a certain
limiting reduced temperature,
\begin{equation}
 T\equiv\frac{1}{\beta rJ}=\frac{1}{4},
\end{equation}
for all values of $\Delta$. The eigenvalues of the recursion relations
\eqref{MF} about this trivial fixed point are a pair of complex conjugates,
\begin{equation}
 \lambda_{\pm}=\frac{\beta Jr}{4}\left(  \cos2\Delta\pm i\sin2\Delta\right)  ,
\end{equation}
which is an indication of the transition to a modulated structure. 

In particular, for the case $\Delta=0$, which corresponds to absence of chirality, 
recursion relations (\ref{MF}) are reduced to simple expressions, 
\begin{equation}%
\begin{split}
\rho_{1}^{\prime} &  =\frac{\exp\left(  c\rho_{1}\right)  }{\exp\left(
c\rho_{1}\right)  +\exp\left(  c\rho_{2}\right)  +\exp\left[  c\left(
1-\rho_{1}-\rho_{2}\right)  \right]  },\\
& \\
\rho_{2}^{\prime} &  =\frac{\exp\left(  c\rho_{2}\right)  }{\exp\left(
c\rho_{1}\right)  +\exp\left(  c\rho_{2}\right)  +\exp\left[  c\left(
1-\rho_{1}-\rho_{2}\right)  \right]  },
\end{split}
\end{equation}
where $c=3\beta Jr/4$. As a result, one can check that the disordered fixed point, 
$\rho_{1}^{\ast}=\rho_{2}^{\ast}=1/3$, is stable at high temperatures, $\beta Jr<4$. 
There is also an ordered fixed point, $\rho_{1}^{\ast}=\rho_{2}^{\ast}=\rho^{\ast}\neq1/3$, 
which comes from the equation
\begin{equation}
\frac{9}{4}\beta Jrw=\ln\frac{1+6w}{1-3w},\label{potts3}%
\end{equation}
where%
\begin{equation}
\rho^{\ast}=\frac{1}{3}-w,
\end{equation}
which is the standard mean-field equation for the three-state Potts model. However, it is 
important to mention that there is always a disordered solution, $w=0$. At low temperatures, 
for $\beta Jr<4$, it is also possible to notice that there is another (ordered) solution, $w\neq0$. 
However, a careful numerical analysis of \eqref{potts3} shows that the ordered solution is already present in
a small range of temperatures $T>1/4$, which is a typicalfeature of first-order transitions.

\begin{figure} [h]
 \centering
 \includegraphics[scale=0.6]{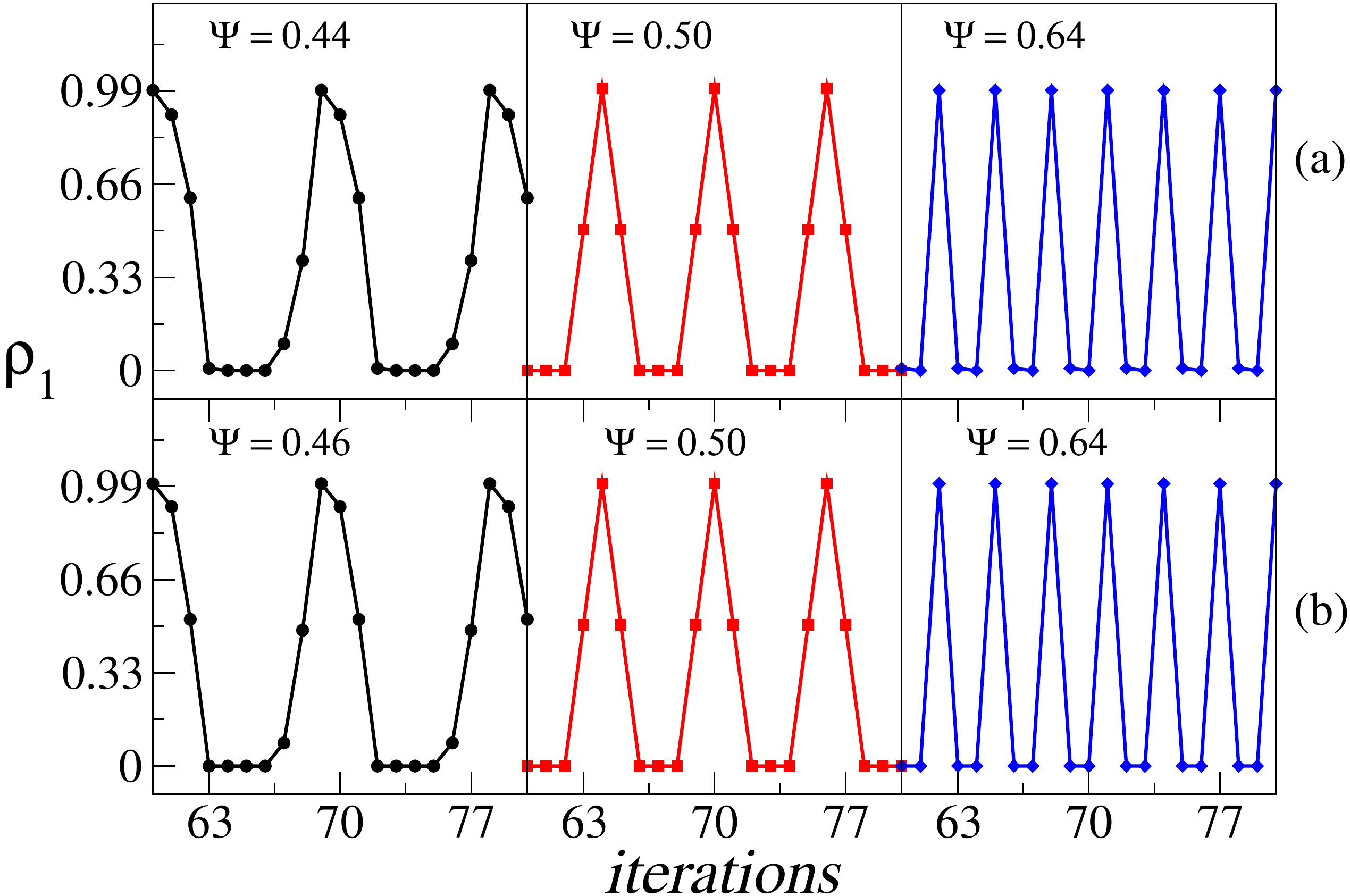}
 \caption{Examples of the behavior of $\rho_1$ in periodic solutions 
  (the behavior of $\rho_2$ is   identical up to a horizontal shift) 
   for (a) $T=0.05$ and (b) $T=0.03$. The values of dimensionless chiral 
   field $\Psi$ and reduced temperature  $T$ have been chosen to obtain short periods.}
\label{periodical}
\end{figure}
In order to study the phase behavior of the model, we perform extensive numerical investigations of the MF equations 
\eqref{MF}, which we may write as 
\begin{equation} \label{MFIt}
 \begin{split}
  \rho_1^{\prime} &= f_1\left(\rho_1,\rho_2\right), \\
  \rho_2^{\prime} &= f_2\left(\rho_1,\rho_2\right),
 \end{split}  
\end{equation}
Such set of coupled nonlinear equations can be solved by means of iterative procedures. 
It is worth to mention that due to the Cayley tree structure, MF equations express a 
relation between densities $\rho_1$ and $\rho_2$ along the generations. In practice, 
we have a nonlinear mapping problem, where primed densities $\rho_1^{\prime}$ and 
$\rho_2^{\prime}$, associated with some generation, which we call $s$, are related to 
unprimed densities $\rho_1$ and $\rho_2$ of generation $s-1$. Then, for fixed $T$ and 
$\Psi$ given by \eqref{ChiralField}, we start with different initial conditions and let 
iterate the map \eqref{MFIt} until it reaches some kind of stationary  behavior whose 
features determine the character of the nature of the phase. As for usual  maps, e.g. 
those describing dynamical systems (see e.g. \cite{cencini}), we find many different 
types of solutions, specifically fixed points and periodic orbits.

As it is sketched above, for $\Psi=0$ there is a disordered, or isotropic,
phase $\rho_{1}=\rho_{2}=1/3$. We shall see below that this phase appears at
sufficiently high temperature, even for $\Psi \neq 0$. Similarly, stationary
fixed points with different densities, $\rho_{1}\neq1/3$ and $\rho_{2}\neq
1/3$, characterize uniform nematic states and are found for values of $\Psi$
and $T$ in certain intervals. A different kind of stationary behavior is
represented by periodic solutions, which appear for some range of parameters.
In this case, $\rho_{1}$ and $\rho_{2}$ do not attain stationary values but
change periodically. This corresponds to situations in which, instead of
reaching a simple fixed point, $\left(  \rho_{1}^{\ast},\rho_{2}^{\ast
}\right)$, the map \eqref{MFIt} flows to a more complex attractor. If the $n$th
iterate of the map flows to a fixed point, $f_{i}^{\left(  n\right)  }%
=f_{i}\left(  f_{i}\left(  ...f_{i}\left(  \rho_{1},\rho_{2}\right)  \right)
\right)$, we have a solution of period $n$, as it is shown in Figure \ref{periodical}.
periods can be obtained by varying the dimensionless chiral field $\Psi$ and the reduced 
temperature $T$. These solutions can be interpreted as modulated phases, in which the typical angle changes
along the tree from one generation to the next, giving room to genuine chiral phases.

\begin{figure} [h]
 \centering
 \includegraphics[scale=0.5]{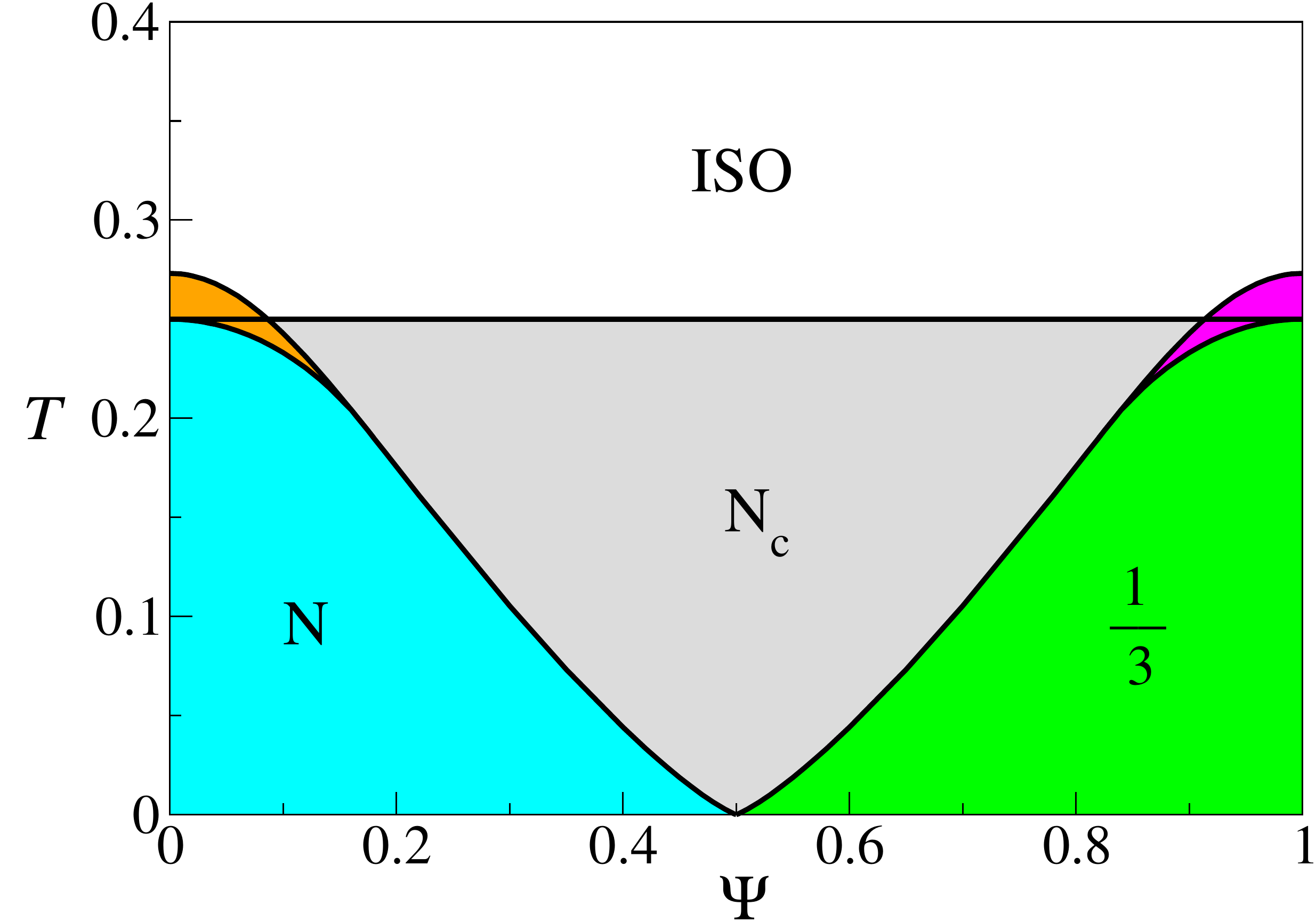}
 \caption{Phase diagram of the chiral nematic clock model with restricted orientations; $T$ is the reduced temperature and 
  $\Psi$ is the dimensionless chiral field. We
  indicate an isotropic phase (ISO), a nematic phase (N$_c$), and a modulated
  nematic phase with wave number $q=2\pi/3$, which we call $1/3$. Region N$_{c}$ is occupied by a
  multitude of chiral nematic phases. Orange and magenta regions present distinct 
  attractors associated with nearby phases.}
 \label{MFPD}
\end{figure}

The main features of the global phase diagram of the model are sketched in Figure \ref{MFPD}, where $T$ 
is the reduced temperature and $\Psi$ is the dimensionless chiral field given by \eqref{ChiralField}. It is 
important to bear in mind that we have a lattice model on a Cayley tree, which means that phase boundaries 
are in fact the stability limit of phases. Also, due to symmetry properties of the model \eqref{Sym}, 
the phase boundaries are symmetrical with respect to the dimensionless chiral field $\Psi=1/2$. In fact, we 
only need to study the system for $0 \le \Psi \le 1/2 $ and map the MF results into $1/2 \le \Psi \le 1  $ 
through \eqref{Sym} and \eqref{ChiralField}. We can identify the region of stability of the nematic phase, 
which is favoured for $0 \le \Psi <1/2$ and sufficient low as well intermediate temperatures. Otherwise, a 
nematic chiral structure, with modulation $q=2\pi/3$, is stable for $1/2 \le \Psi < 1$, but its stability region 
tends to shrink as temperature increases. Nevertheless, for chiral fields in the vicinity of the multiphase point, 
$\Psi=1/2$, thermal fluctuations lead to a fan-shaped region, with many different stable chiral structures (N$_c$), 
between N and $1/3$ states. As expected, the isotropic phase (ISO) is stable for temperatures $T>1/4$. 

The CNC model is an effectively discrete-state spin system closely related to the Potts model. 
The well known mean-field results for the $3$-state Potts model describe a discontinuous 
transition between disordered and ordered phases \cite{Wu1982}. However, for many discrete 
statistical models defined on Cayley trees, it is very complicated to properly characterize 
the type of transition, even by taking the infinite coordination limit, which corresponds to 
mean-field theory. In spite of the difficulties to deal with specific aspects of transitions, our 
numerical investigations show regions which present distinct types of attractors. These 
overlapping regions are indications of phase coexistence as well metastability, which are 
associated with first-order phase transitions. In Figure \ref{MFPD}, the orange region is 
occupied by N and ISO states, for temperatures $T>1/4$, but presents attractors consistent 
with N and N$_c$, for $T<1/4$. Otherwise, in the magenta region, we find  N and $1/3$ solutions 
for $T>1/4$, but it is occupied by $1/3$ and 
N$_c$ for temperatures $T<1/4$.   

\begin{figure} [h]
 \centering
 \includegraphics[scale=0.6]{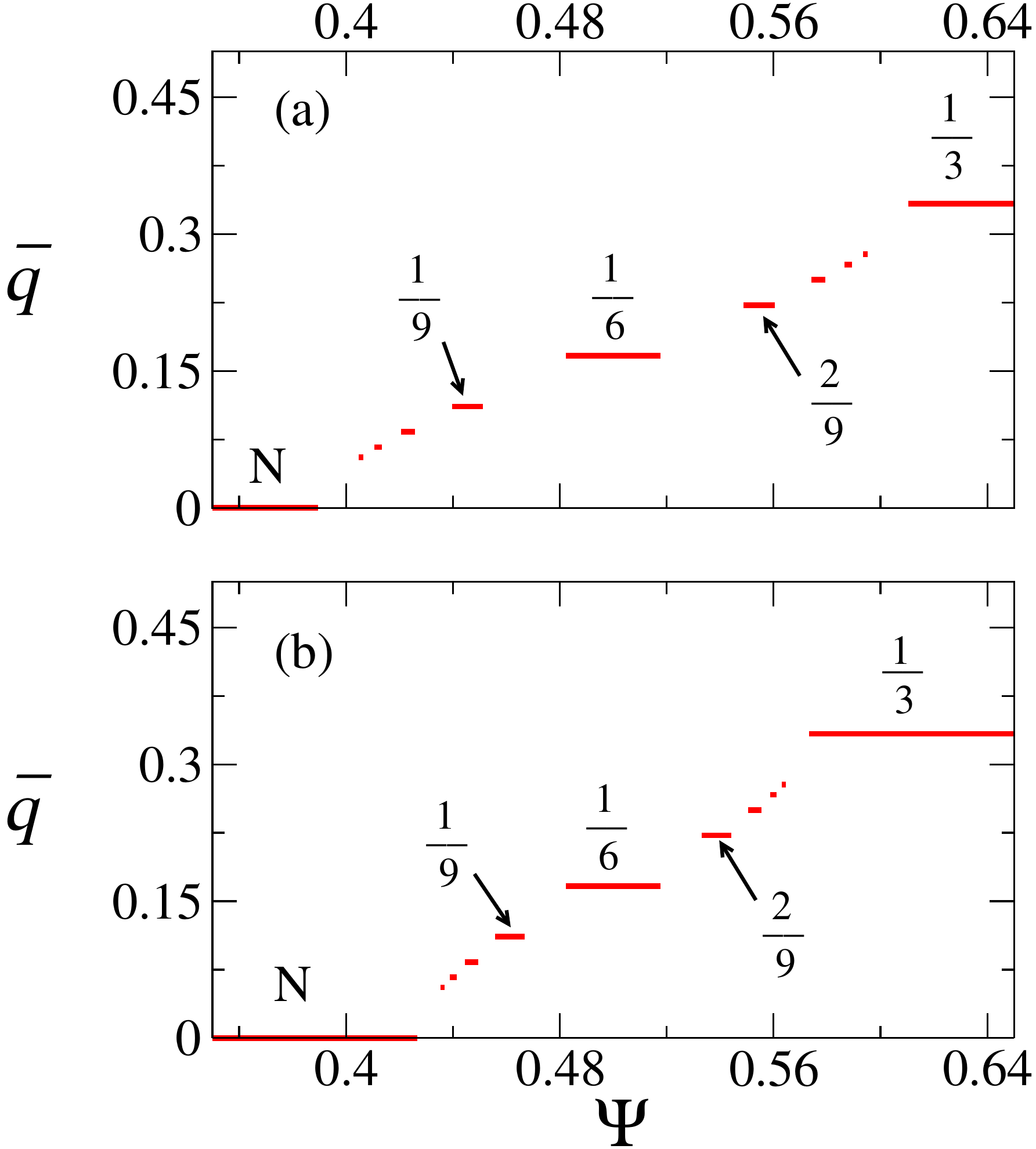}
 \caption{Dimensionless wavenumber $\overline{q}$ as a function of the dimensionless chiral field $\Psi$ for 
  reduced temperatures (a) $T=0.05$ and (b) $T=0.03$. Modulation locks-in some commensurate, nematic-like, 
  chiral phases. Cholesteric pitch exhibits a staircase behavior. }
 \label{Stair}
\end{figure}

The green region occupied by chiral phases is strongly influenced by thermal fluctuations. As a result, 
the modulation presents a nontrivial behavior as a function of temperature and chiral field. Figure 
\ref{Stair} shows the dimensionless wavenumber $\overline{q}=q/2\pi$ as a function of chiral field $\Psi$ for 
different values of temperature. Notice that, at low-temperature regimes, the modulation locks-in at some rational 
values of the type $\overline{q}=L/M$, such as the densities $\rho_1$ and $\rho_2$, which are related to the 
order parameter, evolves along the generations performing a periodic structure with $L$ turns in $M$ iterations. 
Although we only show phases with small values of $M$, we emphasize that long-periodic modulated structures 
are also presented, specially as $T$ increases. In general, one can notice that, for a given temperature, 
as the chiral field increases, N melts into the modulated region, which presents a multitude of modulated phases, 
and a sequence of transitions are observed, until the $1/3$ state is reached. It is possible to show a similar 
behavior (emergence of successive chiral structures) if one considers variations of $T$ for constant $\Psi$ values. 
Note that we only indicate some commensurate phases, but incommensurate states may be found at intermediate 
temperatures, as well as a nontrivial branching mechanism associated with the regions occupied by chiral, 
nematic-like, phases with different modulations \cite{Selke88}.

An analogous staircase-like behavior is presented in the ANNNI and CC models. However, even more 
sophisticated phase sequences are also observed, where the modulation, as a function of model parameters, 
exhibits a fractal structure, which is called devil's staircase \cite{Bak1986,SelkeBak1987,Nascimento2014}. 
Therefore, we argue that the nematic clock model with chiral interactions exhibits nematic chiral phases with 
temperature-dependent pitch. In fact, we do believe the CNC model may also present a case of devil's staircase 
modulation behavior. These peculiar staircases are characterized by a Hausdorff dimension which depends on temperature. 
For the CNC system, it should be interesting, as a further investigation, the study of the fractal character associated 
with the cholesteric pitch.

\section{Conclusions} \label{Con}

We have shown that the Lebwohl-Lasher model used to describe the
nematic-isotropic transition in liquid crystals, with the restriction of the
microscopic nematic directors to point along $p$ planar directions, gives
rise to a nematic counterpart of a $p$-state clock model. We then introduced
chiral interactions between first-neighbor sites along a lattice direction,
according to a mechanism of chirality that has been used to explain
different forms of helimagnetism. The resulting $XY$ chiral Lebwohl-Lasher
statistical model, which we call CNC model, is the nematic counterpart of
the well-investigated (magnetic) chiral clock models. 

On the basis of previous calculations for the analogous magnetic models, we formulated the
statistical problem as a non linear map along the branches of a Cayley
tree. It is known that attractors of this map correspond to solutions of
physical interest. Also, in a suitable limit of infinite coordination of the
tree, equations are somewhat simplified, and we are supposed to recover
standard mean-field results. We take full advantage of the Cayley tree to
investigate the presence of stable modulated structures. 

In this preliminary work, we performed some explicit calculations for a restricted-orientation chiral clock
model with head-tail symmetry. Besides the isotropic and nematic phases, we have shown the 
existence of sequences of modulated structures, with a pitch that depends on temperature 
and a parameter of chirality. A global phase diagram is presented, where it is possible to 
identify uniform ordered states as well spatially modulated phases. These sketchy calculations,
however, are sufficient to illustrate the origins and the main qualitative features of modulation 
associated with helical nematics (cholesteric states) in liquid-crystalline systems. We hope 
this investigation will be a guide for future (and more extensive)work, including connections 
with earlier calculations for statistical molecular models \cite{Krutzen1989,vanderMeer1976,Lin-Liu1976} 
and predictions of the phenomenological elastic theories \cite{Kamien2001}.

\section*{Acknowledgement}

We thank the anonymous referees for the comments and suggestions. 
This work is part of the research program of the INCT-FCx, which is funded by
the Brazilian organizations CNPq and Fapesp. E. S. Nascimento thanks CAPES 
for the financial support. A. Petri is grateful for the kind hospitality of IFUSP. 
E. S. Nascimento is indebted to Walter Selke for critically reading the manuscript.

\appendix
\section{$4$-state nematic chiral clock model} \label{4state}

We now use a similar approach to analyze a four-state $XY$ chiral
Lebwohl-Lasher model. We consider just two traceless tensor microscopic
states,%
\begin{equation}
\mathbf{Q}\left(  1\right)  =\frac{1}{2}\left(
\begin{array}
[c]{cc}%
1 & 0\\
0 & -1
\end{array}
\right)  ,\quad\mathbf{Q}\left(  2\right)  =\frac{1}{2}\left(
\begin{array}
[c]{cc}%
-1 & 0\\
0 & 1
\end{array}
\right)  ,
\end{equation}
from which we write the recursion relations%
\begin{equation}%
\begin{split}
Z_{1}^{\prime} &  =\left[  Z_{1}\exp\left(  \frac{1}{2}\beta J\cos
2\Delta\right)  +Z_{2}\exp\left(  -\frac{1}{2}\beta J\cos2\Delta\right)
\right]  ^{r}\\
Z_{2}^{\prime} &  =\left[  Z_{1}\exp\left(  -\frac{1}{2}\beta J\cos
2\Delta\right)  +Z_{2}\exp\left(  \frac{1}{2}\beta J\cos2\Delta\right)
\right]  ^{r}.
\end{split}
\end{equation}
We then introduce the density variables
\begin{equation}
\rho_{i}=\frac{Z_{i}}{Z_{1}+Z_{2}},
\end{equation}
for $i=1,2$, so that the problem is reduced to a one-dimensional map,%
\begin{equation}
\rho_{1}^{\prime}=\frac{1}{D_{4}}\left[  \exp\left(  -\frac{1}{2}\beta
J\cos2\Delta\right)  +2\rho_{1}\sinh\left(  \frac{1}{2}\beta J\cos
2\Delta\right)  \right]  ^{r},
\end{equation}
with%
\begin{equation}
D_{4}=\left[  \exp\left(  -\frac{1}{2}\beta J\cos2\Delta\right)  +2\rho
_{1}\sinh\left(  \frac{1}{2}\beta J\cos2\Delta\right)  \right]  ^{r}+
\end{equation}%
\[
+\left[  \exp\left(  \frac{1}{2}\beta J\cos2\Delta\right)  -2\rho_{1}%
\sinh\left(  \frac{1}{2}\beta J\cos2\Delta\right)  \right]  ^{r}.
\]
Taking into account that%
\begin{equation}
Q=\left\langle Q_{11}\right\rangle =\frac{1}{2}\rho_{1}-\frac{1}{2}\rho
_{2}=\rho_{1}-\frac{1}{2},
\end{equation}
this map can be written as $Q^{\prime}=f\left(  Q\right)  $.

In the limit of infinite coordination ($r\rightarrow\infty$, $J\rightarrow0$,
with fixed values of $rJ$), we have the simple form%
\begin{equation}
Q^{\prime}=\frac{1}{2}\tanh\left[  \beta Jr\cos\left(  2\Delta\right)
Q\right]  ,
\end{equation}
from which we see that the disordered fixed point is linearly stable for%
\begin{equation}
\frac{k_{B}T}{rJ}>\left\vert \frac{1}{2}\cos2\Delta\right\vert .
\end{equation}
In the $T-\Delta$ phase diagram, besides a trivial ferromagnetic structures,
there is only a quite trivial antiferromagnetic arrangement.

\end{document}